\documentclass[aps, pra, showpacs,superscriptaddress , unsortedaddress, twocolumn]{revtex4-1}
\usepackage{graphicx}
\usepackage{calc}
\usepackage{amsmath}
\usepackage{amssymb}

\begin{document}

\title{Elastic interaction between colloidal particles in confined nematic liquid crystals}

\author{S. B. Chernyshuk }

\affiliation{ Institute of Physics, NAS Ukraine, Prospekt
Nauki 46, Kyiv 03650, Ukraine}

\affiliation{ Bogolyubov Institute of
Theoretical Physics,NAS Ukraine,Metrologichna 14-b, Kyiv 03680,
Ukraine.}

\author{B.I.Lev}

\date{\today}

\begin{abstract}
The theory of elastic interaction of micron size axially symmetric colloidal particles immersed into confined nematic liquid crystal has been
proposed. General formulas are obtained for the self energy of one colloidal particle and interaction energy between two particles in arbitrary confined NLC with strong anchoring condition on the bounding surface.  Particular cases of dipole-dipole interaction in the homeotropic and planar nematic cell with thickness $L$ are considered and found to be exponentially screened on far distances with decay length $\lambda_{dd}=\frac{L}{\pi}$. It is predicted that bounding surfaces in the planar cell crucially change the attraction and repulsion zones of usual dipole-dipole interaction.  As well it is predicted that \textit{the decay length} in quadrupolar interaction is \textit{two times smaller} than for the dipolar case.
\end{abstract}

\maketitle

Colloidal particles in nematic liquid crystals (NLC) have attracted a great research interest during the last years. Anisotropic properties of the host fluid - liquid crystal give rise to a new class of colloidal anisotropic interactions that never occurs in isotropic hosts. The anisotropic interactions result in different
structures of colloidal particles such as linear chains in inverted nematic emulsions \cite{po1,po2}, 2D crystals \cite{Mus}
and 2D hexagonal structures at nematic-air interface
\cite{nych,R10}.

Study of anisotropic colloidal interactions has been made both experimentally \cite{po2}-\cite{jap} and theoretically \cite{lupe}-\cite{perg2}. The first theoretical approach was developed in \cite{po1,lupe} with help of ansatz functions for the director and using multiple expansion in the far field area.
 Another approach \cite{lev,lev3} gave possibility to find approximate solutions in terms of geometrical shape of particles. Recently authors of \cite{perg1,perg2} proposed a method for finding elastic interaction between colloids based on the fixing of director field on the surface of virtual sphere surrounding the real particle. The predicted dipole-dipole forces are three times weaker and quadrupole-quadrupole five times weaker than results of \cite{lupe}. On the other hand authors of \cite{jap} recently have measured experimentally both interactions and found that experimental results are in accordance with Lubensky \textit{et al.} prediction \cite{lupe} with about $10\%$ accuracy. This allows to justify assumptions of \cite{lupe} for spherical particles for intinite nematic liquid crystal. In this paper we suggest to generalize that approach for the case of the confined nematic liquid crystals as practically always NLC has to be confined with walls, cells or containers. In a broader context, understanding the elasticity-mediated colloidal interactions in confined media is of great importance not only in the field of regular thermotropic liquid crystals, but also for understanding interactions in more complex media with orientational order, for example, in solutions of DNA, f-actin and other biologically relevant molecules. Up to now almost all experimental studies did not take into account quantitatively confinement effects besides the article of Vilfan \textit{et al.} \cite{conf}. In that paper authors have found exponential screening effects for quadrupole-quadrupole interaction between spherical particles in homeotropic NLC cell.  From our viewpoint there was only one theoretical approach for description of colloidal particles in confined NLC performed in papers \cite{fukuda1,fukuda}. 

In this paper we propose the new approach for quantitative description of the axial colloidal particles in confined NLC. This method enables to find self energy of one colloidal particle and interaction energy between two particles in arbitrary confined NLC with strong anchoring condition $n_{\mu}(\textbf{s})=0$ on the bounding surface. We apply general formulas to the particular cases of dipole-dipole and quadrupole-quadrupole interaction in the homeotropic cell and to the dipole-dipole interaction in the planar cell with thickness $L$. 

Consider axially symmetric particle of the size $0.1 \mu m \div 10 \mu m$ which may carry topological defects such as hyperbolic hedgehog, diclination ring or boojums. Director field far from the particle in the infinite LC has the form $n_{x}(\textbf{r})=p\frac{x}{R^{3}}+3c\frac{xz}{R^{5}},n_{y}(\textbf{r})=p\frac{y}{R^{3}}+3c\frac{yz}{R^{5}}$ with $p$ and $c$ being dipole and quadrupole moment (we use another notation for $c$ with respect to the $\tilde{c}$ in \cite{lupe}, so that our $c=\frac{2}{3}\tilde{c}$ ). It was found in \cite{lupe} that $p=\alpha a^2$,$c=-\beta a^3$ with $a$ being the particle radius, and for instance $\alpha=2.04$,$\beta=0.72$ for hyperbolic hedgehog configuration. In order to find energy of the system: particle(s) + LC it is necessary to introduce some effective functional $F_{eff}$ so that it's Euler-Lagrange equations should have the above solutions. In the \cite{lupe} it was found that in the one constant approximation with Frank constant $K$ the effective functional has the form:
\begin{widetext}
\begin{equation}
F_{eff}=K\int d^{3}x\left\{\frac{(\nabla n_{\mu})^{2}}{2}-4\pi P(\textbf{x})\partial_{\mu}n_{\mu}-4\pi C(\textbf{x})\partial_{z}\partial_{\mu}n_{\mu} \right\}\label{flin}
\end{equation}
\end{widetext}
which brings Euler-Lagrange equations:
\begin{equation}
\Delta n_{\mu}=4\pi\left[\partial_{\mu}P(\textbf{x})-\partial_{z}\partial_{\mu}C(\textbf{x})\right] \label{nmu}
\end{equation}
where $P(\textbf{x})$ and $C(\textbf{x})$ are dipole- and quadrupole moment densities.
For the infinite space the solution has the known form: $n_{\mu}(\textbf{x})=\int d^{3}\textbf{x}' \frac{1}{\left|\textbf{x}-\textbf{x}'\right|}\left[ -\partial_{\mu}'P(\textbf{x}')+\partial_{\mu}'\partial_{z}'C(\textbf{x}') \right] $. If we consider $P(\textbf{x})=p\delta(\textbf{x})$ and $C(\textbf{x})=c\delta(\textbf{x})$ this really brings $n_{x}(\textbf{r})=p\frac{x}{R^{3}}+3c\frac{xz}{R^{5}},n_{y}(\textbf{r})=p\frac{y}{R^{3}}+3c\frac{yz}{R^{5}}$.

In the case of confined nematic with the boundary conditions $n_{\mu}(\textbf{s})=0$ on the surface $S$
the solution of EL equation has the form:
\begin{equation}
n_{\mu}(\textbf{x})=\int_{V} d^{3}\textbf{x}' G(\textbf{x},\textbf{x}')\left[ -\partial_{\mu}'P(\textbf{x}')+\partial_{\mu}'\partial_{z}'C(\textbf{x}') \right] \label{solmain}
\end{equation}
where $G$ is the Green function $\Delta_{\textbf{x}}G(\textbf{x},\textbf{x}')=-4\pi \delta(\textbf{x}-\textbf{x}')$ for  $\textbf{x},\textbf{x}'\in \textbf{V}$ and $G(\textbf{x},\textbf{s})=0 $ for any \textbf{s} of the bounding surfaces. Consider $N$ particles in the confined NLC, so  $P(\textbf{x})=\sum_{i}p_{i}\delta(\textbf{x}-\textbf{x}_{i})$ and $C(\textbf{x})=\sum_{i}c_{i}\delta(\textbf{x}-\textbf{x}_{i})$. Then substition (\ref{solmain}) into $F_{eff}$ brings: $F_{eff}=U^{self}+U^{interaction}$
where  $ U^{self}=\sum_{i}U_{i}^{self} $ , here $U_{i}^{self}$ is the interaction of the $i$-th particle with the bounding surfaces $ U_{i}^{self}=U_{dd}^{self}+U_{dQ}^{self}+U_{QQ}^{self} $. In general case the interaction of the particle with bounding surfaces (self-energy part) takes the form:  
 $$
 U_{dd}^{self}=-2\pi K p^{2}\partial_{\mu}\partial_{\mu}'H(\textbf{x}_{i},\textbf{x}_{i}')|_{\textbf{x}_{i}=\textbf{x}_{i}'}
 $$
 \begin{equation}
 U_{dQ}^{self}=-4\pi K pc \partial_{\mu}\partial_{\mu}'\partial_{z}'H(\textbf{x}_{i},\textbf{x}_{i}')|_{\textbf{x}_{i}=\textbf{x}_{i}'} \label{uself}
 \end{equation}
 $$
 U_{QQ}^{self}=-2\pi K c^{2} \partial_{z}\partial_{z}'\partial_{\mu}\partial_{\mu}'H(\textbf{x}_{i},\textbf{x}_{i}')|_{\textbf{x}_{i}=\textbf{x}_{i}'}
 $$
where  $ G(\textbf{x},\textbf{x}')=\frac{1}{|\textbf{x}-\textbf{x}'|}+H(\textbf{x},\textbf{x}') $ and  $\Delta_{\textbf{x}}H(\textbf{x},\textbf{x}')=0$ (we excluded divergent part of self energy from $\frac{1}{|\textbf{x}-\textbf{x}'|}$).\\
Interaction energy $ U^{interaction}=\sum_{i<j}U_{ij}^{int} $.  Here $U_{ij}^{int}$ is the interaction energy between $i$ and $j$ particles: 
 $ U_{ij}^{int}= U_{dd}+U_{dQ}+U_{QQ} $:
\begin{equation}
U_{dd}=-4\pi K pp'\partial_{\mu}\partial_{\mu}'G(\textbf{x}_{i},\textbf{x}_{j}') \label{uint}
\end{equation}

$$
U_{dQ}=-4\pi K \left\{pc'\partial_{\mu}\partial_{\mu}'\partial_{z}'G(\textbf{x}_{i},\textbf{x}_{j}')+p'c\partial_{\mu}'\partial_{\mu}\partial_{z}G(\textbf{x}_{i},\textbf{x}_{j}')\right\}
$$

$$
U_{QQ}=-4\pi K cc'\partial_{z}\partial_{z}'\partial_{\mu}\partial_{\mu}'G(\textbf{x}_{i},\textbf{x}_{j}')
$$
Here unprimed quantities are used for particle $i$ and primed for particle $j$.
Formulas (\ref{uself}) and (\ref{uint}) represent general expressions for the self energy of one particle (energy of interaction with the walls) and interparticle elastic interactions in the arbitrary confined NLC with strong anchoring conditions $n_{\mu}(\textbf{s})=0$ on the bounding surfaces. Below we will apply these expressions for particular cases of the nematic cell with homeotropic and planar configurations. 

\subsection{Interaction in the homeotropic cell with width $L$}
Green function in this case has the form \cite{jac} :
\begin{widetext}
\begin{equation}
G_{hom}^{cell}(\textbf{x},\textbf{x}')=\frac{4}{L}\cdot\sum_{n=1}^{\infty}\sum_{m=-\infty}^{\infty}e^{im({\varphi-\varphi'})}sin\frac{n\pi z}{L}sin\frac{n\pi z'}{L}I_{m}(\frac{n\pi \rho_{<}}{L})K_{m}(\frac{n\pi \rho_{>}}{L})
\end{equation}
\end{widetext}
Here heights $z,z'$, horizontal projections $\rho_{<},\rho_{>}$ and $I_{m},K_{m}$ are modified Bessel functions.
Then using of (\ref{uint}) brings dipole-dipole interaction in the cell :
\begin{equation}
U_{dd,hom}^{c}=\frac{16\pi K pp'}{L^{3}}\sum_{n=1}^{\infty}(n\pi)^{2} sin\frac{n\pi z}{L}sin\frac{n\pi z'}{L}K_{0}(\frac{n\pi \rho}{L})\label{dhom}
\end{equation}
with $\rho$ being the horizontal projection of the distance between particles. Similar
quadrupole-quadrupole interaction takes the form:
\begin{equation}
U_{QQ,hom}^{c}=\frac{16\pi K cc'}{L^{5}}\sum_{n=1}^{\infty}(n\pi)^{4}cos\frac{n\pi z}{L}cos\frac{n\pi z'}{L}K_{0}(\frac{n\pi \rho}{L})
\end{equation}

\begin{figure}[ht!]
\includegraphics[clip=,width=\linewidth]{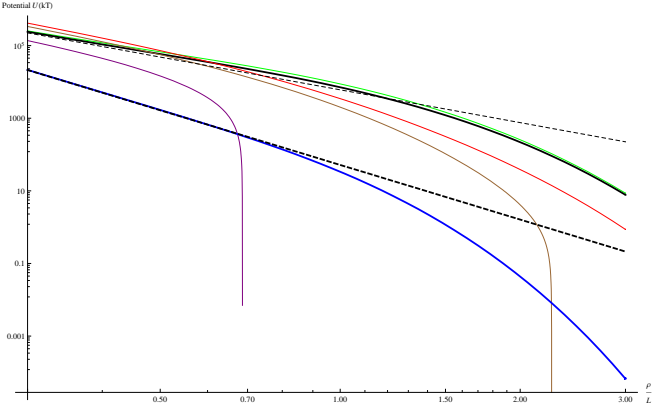}
\caption{\label{Fig1}(Color online) Log-log plots of the interaction potential in $kT$ units as a function of the rescaled interparticle distance $\rho/L$. Here particle's radius $a=2.2\mu m$,cell thickness $L=7\mu m$,$K=7 pN$,$p=p'=2.04a^{2}$,$c=c'=0.2a^{3}$. Blue thick line is quadrupole potential in homeotropic cell (it approximates experimental data from \cite{conf} very well), dashed thick line is it's power-law asymptotics $\propto1/\rho^{5}$. All thin lines are dipole potentials in planar cell. Coming anticlockwise first purple line is \textit{attraction} along the direction $\varphi=40^{o}$, second brown line is along $\varphi=20^{o}$, third red line is along $z$ axis $\varphi=0$ and the last green thin line is \textit{repulsion} along $\varphi=\frac{\pi}{2}$. Black thick line is dipole repulsion in homeotropic cell from (\ref{dhom}). Thin dashed line is the power-law asymptotics $U=\frac{4\pi K p^{2}}{\rho^{3}}$.  }
\end{figure}
\begin{figure}[ht!]
\includegraphics[clip=,width=\linewidth,]{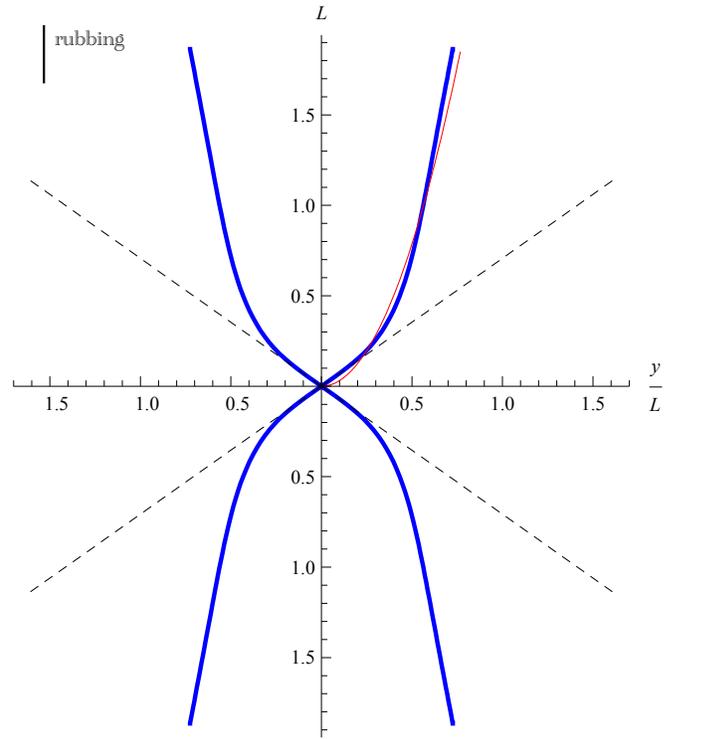}
\caption{\label{Fig2}(Color online) Blue thick line is the border of the attraction (inside) and repulsion (outside) zone for parallel dipoles in the planar cell from (\ref{uplanfi}). Director $\textbf{n}_{0}||z$. Red thin line is the parabola $z=\frac{\pi y^{2}}{L}$. Dashed lines make angle $\varphi=arccos(\frac{1}{\sqrt{3}})$ with $z$ and are borders of repulsion and attraction zone for unlimited nematic.}
\end{figure}
When both particles are located in the center of the cell $z=z'=\frac{L}{2}$ we have $U_{dd,hom}^{c}=\frac{16\pi K pp'}{L^{3}}\sum_{n=1,odd}^{\infty}(n\pi)^{2}K_{0}(\frac{n\pi \rho}{L})$ (see black thick line on the Fig.~\ref{Fig1}). In the limit of small distance $\rho \ll L$ between particles it has  asymptotic $U_{dd,hom}^{c} \rightarrow \frac{4\pi K pp'}{\rho^{3}}$ that is in agreement with standard formula for the usual dipole-dipole interaction $U_{dd}=\frac{4\pi K pp'}{r^{3}}(1-3cos^2\theta)$ for $\theta =\frac{\pi}{2}$. From the Fig.~\ref{Fig1} it is clearly seen that power-law behaviour $U\propto \frac{1}{\rho^{3}}$ is valid to the $\rho=1.2L$. For larger distances $\rho>1.2L$ exponential decay takes place $U_{dd,r\Rightarrow \infty}=16\pi K pp'\frac{\pi^{2}}{L^{2}}\frac{e^{-\frac{\pi \rho}{L}}}{\sqrt{2L\rho}}$ so
we have decay length for dipole-dipole interaction $\lambda_{dd}=\frac{L}{\pi}$. \\
When both particles are located in the center of the cell $z=z'=\frac{L}{2}$ we have quadrupole interaction $U_{QQ,hom}^{c}=\frac{16\pi K cc'}{L^{5}}\sum_{n=2,even}^{\infty}(n\pi)^{4}K_{0}(\frac{n\pi \rho}{L})$(see thick blue line on the Fig.~\ref{Fig1}). This coinsides with the result of \cite{fukuda} if we take $\Gamma$ there to be equal $\Gamma=2\pi Kc=-2\beta\pi Ka^{3}$. Let's emphasize that in \cite{fukuda} the $\Gamma$ remains unknown quantity.
In the limit of small distance $\rho \ll L$ between particles it has  asymptotics $U_{QQ,hom}^{c} \rightarrow \frac{36\pi K cc'}{\rho^{5}}$ that is in agreement with standard formula for the usual quadrupole-quadrupole interaction $U_{QQ}=4\pi K cc'\frac{9-90cos^{2}\theta+105cos^{4}\theta}{r^{5}}$ for $\theta =\frac{\pi}{2}$. This power-lar behaviour is valid to the distance $\rho=0.8L$.
For larger distance $\rho>0.8L$ crossover to the exponential decay occurs  $U_{QQ,r\Rightarrow \infty}=8\pi K cc'(\frac{2\pi}{L})^{4}\frac{e^{-\frac{2\pi \rho}{L}}}{\sqrt{L\rho}}$. So we come to the following prediction:
\textit{decay length} for quadrupole particles:$\lambda_{QQ}=\frac{L}{2\pi}=\frac{\lambda_{dd}}{2}$ \textit{is twice smaller} than for dipole particles.

\subsection{Interaction in the planar cell with thickness $L$}
In order to find Green function for this case let's turn coordinate system (CS) of the homeotropic cell $CS^{hom}$ $(x,y,z)$ round the $y$ axis on $\pi/2$. Then we will have $CS^{plan}$ $(\tilde{x},\tilde{y},\tilde{z})$ with transition matrix $A$: $\textbf{x}=A\tilde{\textbf{x}},\textbf{x}'=A\tilde{\textbf{x}}'$ so that $x=\tilde{z},y=\tilde{y},z=-\tilde{x}$. Then $G_{hom}(\textbf{x},\textbf{x}')=G_{hom}(A\tilde{\textbf{x}},A\tilde{\textbf{x}}')=G_{plan}(\tilde{\textbf{x}},\tilde{\textbf{x}}')$. Omitting sign $\sim$ we may write Green function for planar cell in the $CS^{plan}$ with $\textbf{n}||z$ and $x$ perpendicular to the cell plane ($x\in [0,L]$):
\begin{widetext}
\begin{equation}
G_{plan}^{cell}(\textbf{x},\textbf{x}')=\frac{4}{L}\cdot\sum_{n=1}^{\infty}\sum_{m=-\infty}^{\infty}e^{im({\varphi-\varphi'})}sin\frac{n\pi x}{L}sin\frac{n\pi x'}{L}I_{m}(\frac{n\pi \rho_{<}}{L})K_{m}(\frac{n\pi \rho_{>}}{L})
\end{equation}
where heights $x,x'$, horizontal projections $\rho_{<}=\sqrt{y^{2}+z^{2}},\rho_{>}=\sqrt{y'^{2}+z'^{2}}$ , $tg \varphi = \frac{y}{z},tg \varphi' = \frac{y'}{z'}$ and $\rho_{<}$ is less than $\rho_{>}$.
Then taking derivatives brings dipole-dipole interaction in the planar cell $U_{dd,plan}^{c}=-4\pi K pp'\partial_{\mu}\partial_{\mu}'G_{plan}^{cell}$:
\begin{equation}
U_{dd,plan}^{c}=\frac{16\pi K pp'}{L^{3}}(F_{1}-F_{2}cos^{2}\varphi)\label{uplanfi}
\end{equation}
where
$$
F_{1}=\sum_{n=1}^{\infty}\frac{(n\pi)^{2}}{2}sin\frac{n\pi x}{L}sin\frac{n\pi x'}{L}\left[K_{0}(\frac{n\pi \rho}{L})+K_{2}(\frac{n\pi \rho}{L})\right]-(n\pi)^{2}cos\frac{n\pi x}{L}cos\frac{n\pi x'}{L}K_{0}(\frac{n\pi \rho}{L}),
$$
$$
F_{2}=\sum_{n=1}^{\infty}(n\pi)^{2}sin\frac{n\pi x}{L}sin\frac{n\pi x'}{L}K_{2}(\frac{n\pi \rho}{L}) .
$$
\end{widetext}
When both particle are located in the center of the cell $x=x'=\frac{L}{2}$ we have $F_{1}=\sum_{n=1,odd}^{\infty}\frac{(n\pi)^{2}}{2}\left[K_{0}(\frac{n\pi \rho}{L})+K_{2}(\frac{n\pi \rho}{L})\right]$ $ - \sum_{n=2,even}^{\infty}(n\pi)^{2}K_{0}(\frac{n\pi \rho}{L})$ and $F_{2}=\sum_{n=1,odd}^{\infty}(n\pi)^{2}K_{2}(\frac{n\pi \rho}{L})$. In the limit of small distance $\rho \ll L$ between particles these functions have asymptotics $F_{1}\rightarrow \frac{L^{3}}{4\rho^{3}}$ and $F_{2}\rightarrow \frac{3L^{3}}{4\rho^{3}}$ so that we come to the well known result $U_{dd}=\frac{4\pi K pp'}{\rho^{3}}(1-3cos^{2}\varphi)$ for $\rho \ll L$. 
In the limit of big distances $\rho \geq L$ we have $\frac{F_{1}}{F_{2}}=1-\frac{L}{\pi \rho}+o(\frac{L}{\rho})$ with accuracy $5\%$ already for $\rho=L$. So for $\rho \geq L$ it may be written $U_{dd,plan}^{c}=\frac{16\pi K pp'}{L^{3}}F_{2}(\rho)\cdot(1-\frac{L}{\pi \rho}-cos^{2}\varphi)$ so that dipole-dipole interaction is attractive for $-\varphi_{c}\leq\varphi<\varphi_{c}$, $\varphi_{c}=arccos(\sqrt{1-\frac{L}{\pi \rho}})\approx \sqrt{\frac{L}{\pi \rho}}$ and is repulsive for $\varphi_{c}<\varphi<2\pi-\varphi_{c}$ (if dipoles are parallel each other $p=p'$ and vice versa if $p=-p'$  ). In other words for $\rho>L$ dipole-dipole interaction is attractive inside parabola $z=\frac{\pi y^{2}}{L}$ and is repulsive outside this parabola (see Fig.~\ref{Fig2}). Practically this parabola $z=\frac{\pi y^{2}}{L}$ with enough eccuracy confines the attraction and repulsive zone even for smaller distances $\rho\geq 0.3 L$ as it is seen from the Fig.~\ref{Fig2}. All numerical calculations in the paper were performed using Mathematica 6, and in all series we used summation $\sum_{n=1}^{300}$.

To conclude we have found general approach for description of the axial colloidal particles of the size $0.1 \mu m \div 10 \mu m$ in the confined NLC. The \textit{decay length} for dipole interaction is found to be \textit{twice more} than for quadrupole interaction in the homeotropic cell. In the planar cell bounding surfaces crucially change attraction and repulsion zones for the distances larger than $\rho_{c}=0.33 L$ where \textit{crossover} to the parabola $z=\frac{\pi y^{2}}{L}$ takes place, so that attraction zone is inside this parabola and repulsive zone is outside it. This approach  has been succesfully applied as well for the interaction of one particle with the one homeotropic and planar wall and for interaction between two particles near such wall. That results will be published in the upcoming paper \cite{we}.

\end{document}